\documentclass[aps,twocolumn,prb,amsmath,amssymb]{revtex4}

\usepackage{graphicx,epsfig}
\usepackage{color}
\usepackage{wasysym}
\usepackage[normalem]{ulem}

\begin{document}
\title{Nonsymmetrized noise in a quantum dot: Interpretation in terms of energy transfer and coherent superposition of scattering paths}
\author{R. Zamoum$^{1,2}$}
\author{M. Lavagna$^{3*}$}
\author{A. Cr\'epieux$^{1}$}
\affiliation{$^1$ Aix Marseille Universit\'e, Universit\'e de Toulon, CNRS, CPT UMR 7332, 13288 Marseille, France}
\affiliation{$^2$ Facult\'e des sciences et des sciences appliqu\'ees, Universit\'e de Bouira, rue Drissi Yahia, Bouira 10000, Algeria}
\affiliation{$^3$ Universit\'e Grenoble Alpes INAC-SPSMS F-38000 Grenoble, France\\
CEA INAC-SPSMS F-38000 Grenoble, France}

\begin{abstract}
We calculate the nonsymmetrized current noise in a quantum dot connected to two reservoirs by using the non-equilibrium Green function technique. We show that both the current auto-correlator (inside a single reservoir) and the current cross-correlator (between the two reservoirs) are expressed in terms of transmission amplitude and coefficient through the barriers. We identify the different energy-transfer processes involved in each contribution to the auto-correlator, and we highlight the fact that when there are several physical processes, the contribution results from a coherent superposition of scattering paths. Varying the gate and bias voltages, we discuss the profile of the differential Fano factor in light of recent experiments, and we identify the conditions for having a distinct value for the auto-correlator in the left and right reservoirs.
\end{abstract}

\maketitle


\section{Introduction}

Since the beginning of the 2000s, it has been known that to characterize finite-frequency current fluctuations in quantum conductors, one has to measure and calculate the nonsymmetrized noise, which corresponds to the emission noise at positive frequency and to the absorption noise at negative frequency~\cite{aguado00,deblock03}. This is related to the fact that the current operators do not commute for quantum systems~\cite{lesovik97,gavish00}. In taking the symmetrized noise, one mixes the emission and absorption parts and the relevant information is lost. Despite the growing interest in nonsymmetrized finite-frequency (NSFF) noise in quantum systems both experimentally~\cite{onac06a,billangeon06,onac06b,zakka07,basset10,basset12,parmentier12,altimiras14,parlavecchio15} and theoretically~\cite{engel04,lebedev05,martin05,creux06,zazunov07,bena07,safi08,rothstein09,moca11,gabdank11,hammer11,safi11,zamoum12,muller13,rothstein14,roussel15}, there is no clear and direct interpretation of the physical processes that 
contribute to the 
noise, even 
for non-interacting systems. Here we show that it makes sense to interpret each contribution to the NSFF auto-correlators in terms of the energy-transfer process, following the ideas developed by several authors. Indeed, it has been shown that the frequency at which the NSFF noise is evaluated corresponds to the energy provided or absorbed by the detector \cite{engel04} or by the electromagnetic environment, and that current fluctuations produce radiation of photons/plasmons in a phase-coherent conductor \cite{beenakker01,gabelli04,beenakker04,gustavsson07,fulga10,zakka10,lebedev10,schneider10,schneider12,lu13,bednorz13,kaasbjerg15,grimsmo16}. To address this issue from a theoretical point of view, one can use either the scattering theory~\cite{buttiker92,blanter00,kogan09,nazarov09} or the non-equilibrium Green function (NEGF) technique~\cite{jauho94,haug07}. Whereas the former method applies to non-interacting systems, the latter can be extended to interacting ones.  However, the NEGF technique has been used so 
far only for the calculation of symmetrized noise \cite{souza08}.

In this article, we present two methods to derive the noise spectrum for a quantum dot (QD) connected to two reservoirs. The first method is based on the NEGF technique and allows us to express the NSFF auto-correlators and cross-correlators in terms of the transmission amplitude and transmission coefficient through the barriers. The second method is based on a detailed analysis of all the different physical processes that contribute to the auto-correlators by emitting energy in one reservoir, paying attention to considering the coherent superposition of scattering paths when more than one process is involved.  The cross-correlators cannot be obtained within the second method, but we check that for the auto-correlators the two methods lead to the same expressions. We benchmark our results with the known results existing in the literature in some given limits, notably in the non-interacting limit when the scattering theory can be used, and we discuss the charge fluctuations in the QD.

The paper is organized as follows: in Sec.~II, we present the system, the model, and the details of the NSFF noise calculation within the NEGF technique, and we discuss the results. In Sec.~III, we present a new method for the determination of the NSFF auto-correlators based on a careful analysis of the various scattering processes that lead to energy transfer. For an Anderson-type transmission amplitude, we identify in Sec.~IV the conditions to get distinct auto-correlators in the left and right reservoirs, and we discuss the differential Fano factor profile. We conclude in Sec.~V.


\section{nonsymmetrized noise}

We consider a QD connected to two reservoirs described by the non-interacting single-level Anderson Hamiltonian $H=H_L+H_R+H_T+H_D$, where $H_\alpha=\sum_{k\in \alpha} \varepsilon_{k} c_{k}^{\dag}c_{k}$ is the Hamiltonian of the left ($\alpha=L$) and right ($\alpha=R$) reservoirs, $H_T=\sum_{\alpha=L,R}\sum_{k\in \alpha}(V_k c_{k}^{\dag} d+h.c.)$ is the transfer Hamiltonian, and $H_D= \varepsilon_0 d^{\dag}d$ is the Hamiltonian of the QD. $c_{k}^{\dag}(d^{\dag})$ and $c_{k}(d)$ are the creation and annihilation operators of one electron in the reservoirs (QD). The quantities $\varepsilon_0$, $\varepsilon_{k}$ and $V_k$ are respectively the QD energy level, the energy of an electron with momentum $k$ in the reservoir and the transfer matrix element between the corresponding states. The spin degree of freedom can be included without any complication.

The objective is to calculate the NSFF noise defined as
\begin{eqnarray}
{S}_{\alpha\beta}(\omega)=\int_{-\infty}^{\infty} \mathcal{S}_{\alpha\beta}(t,0)e^{-i\omega t}dt~,
\end{eqnarray}
where $\mathcal{S}_{\alpha\beta}(t,0)=\langle \Delta \hat{I}_\alpha(t) \Delta \hat{I}_\beta(0) \rangle$ is the current correlator and $\Delta \hat{I}_\alpha(t)=\hat{I}_\alpha(t)-\langle \hat I_\alpha\rangle$ with $\hat{I}_\alpha$, the current operator from the $\alpha$ reservoir to the central region through the $\alpha$ barrier given by:  $\hat I_\alpha=(ei/\hbar)\sum_{k\in\alpha} \big( V_k c_{k}^{\dag} d-V_k^{\ast} d^{\dag} c_{k} \big)$, and $\langle \hat I_\alpha\rangle$ its  average value.

\subsection{Noise calculation using the NEGF technique}

Following Haug and Jauho~\cite{haug07}, we first substitute the expression of the current operator in the correlator and obtain~\cite{note1}
\begin{eqnarray}\label{correlator}
\mathcal{S}_{\alpha\beta}(t,t')&=&
\frac{e^2}{\hbar^2} \sum_{k\in\alpha , k'\in \beta}\Big[V_kV_{k'}G_1^{cd,>}(t,t')\nonumber\\
&&-V_kV_{k'}^*G_2^{cd,>}(t,t')-V_k^*V_{k'}G_3^{cd,>}(t,t')\nonumber\\
&&+V_k^*V_{k'}^*G_4^{cd,>}(t,t') \Big]-\langle \hat I_\alpha \rangle\langle \hat I_\beta\rangle ~,
\end{eqnarray}
where $G_{i}^{cd,>}(t,t')$, with $i=1$--$4$, are the greater components of the two-particle Green functions mixing $c_k$ and $d$ operators defined as
\begin{eqnarray}
G_1^{cd,>}(t,t')&=&-\langle c_k^\dag(t)d(t)c^\dag_{k'}(t')d(t')\rangle~,\\
G_2^{cd,>}(t,t')&=&-\langle c_k^\dag(t)d(t)d^\dag(t')c_{k'}(t')\rangle~,\\
G_3^{cd,>}(t,t')&=&-\langle d^\dag(t)c_k(t)c^\dag_{k'}(t')d(t')\rangle~,\\
G_4^{cd,>}(t,t')&=&-\langle d^\dag(t)c_k(t)d^\dag(t')c_{k'}(t')\rangle~.
\end{eqnarray}
We introduce 
$G_{i}^{cd} (\tau,\tau')$ with $\tau>\tau'$, the contour-ordered (along the Keldysh contour C \cite{keldysh65}) counterparts of $G_{i}^{cd,>}(t,t')$, as well as $\mathcal{S}_{\alpha\beta}(\tau,\tau')$, the contour-ordered counterpart of  $\mathcal{S}_{\alpha\beta}(t,t')$. We then derive and solve the equations of motion for $G^{cd}_i(\tau,\tau')$ in order to express them in terms of: (i) the contour-ordered one-particle Green function for the disconnected reservoirs defined as $g_k(\tau,\tau')=-i\langle T_C c_k(\tau)c^\dag_{k'}(\tau')\rangle_0$, where $T_C$ is the contour-ordering operator, (ii)  the contour-ordered one-particle Green function for the QD connected to the reservoirs, $G(\tau,\tau')=-i\langle T_C d(\tau)d^\dag(\tau')\rangle$, and (iii) the contour-ordered two-particle Green functions for the QD, $G^{(2)}_i(\tau,\tau',\tau_1,\tau_2)$, defined as
\begin{eqnarray}
G_1^{(2)}(\tau,\tau',\tau_1,\tau_2)&=&-\langle T_Cd(\tau)d(\tau')d^\dag(\tau_1)d^\dag(\tau_2)\rangle~,\\
G_2^{(2)}(\tau,\tau',\tau_1,\tau_2)&=&-\langle T_Cd(\tau)d^\dag(\tau')d(\tau_1)d^\dag(\tau_2)\rangle~,\\
G_3^{(2)}(\tau,\tau',\tau_1,\tau_2)&=&-\langle T_Cd^\dag(\tau)d(\tau')d(\tau_1)d^\dag(\tau_2)\rangle~,\\
G_4^{(2)}(\tau,\tau',\tau_1,\tau_2)&=&-\langle T_Cd^\dag(\tau)d^\dag(\tau')d(\tau_1)d(\tau_2)\rangle~.
\end{eqnarray}
After a series of manipulations, we get
\begin{eqnarray}\label{noise_tau}
&&\mathcal{S}_{\alpha\beta}(\tau,\tau')=\delta_{\alpha\beta}\mathcal{\tilde S}_{\alpha}(\tau,\tau')+\frac{e^2}{\hbar^2} \sum_{k\in\alpha,k'\in \beta} \vert V_k V_{k'} \vert^2 \nonumber \\
 &&
\times\iint d\tau_1 d\tau_2\Big[-g_k(\tau_1,\tau)g_{k'}(\tau_2,\tau') G_1^{(2)}(\tau,\tau',\tau_1,\tau_2)\nonumber \\
 &&
+g_k(\tau_2,\tau)g_{k'}(\tau',\tau_1) G_2^{(2)}(\tau,\tau',\tau_1,\tau_2)\nonumber \\
&& -g_k(\tau,\tau_1)g_{k'}(\tau_2,\tau') G_3^{(2)}(\tau,\tau',\tau_1,\tau_2) \nonumber \\
 &&
 -g_k(\tau,\tau_1)g_{k'}(\tau',\tau_2)  G_4^{(2)}(\tau,\tau',\tau_1,\tau_2) \Big]-\langle \hat I_\alpha \rangle\langle \hat I_\beta\rangle ~,\nonumber \\
\end{eqnarray}
where $\mathcal{\tilde S}_{\alpha}(\tau,\tau')=(e^2/\hbar^2) \sum_{k\in \alpha} \vert V_k \vert^2 \big[g_k(\tau',\tau)G(\tau,\tau')+g_k(\tau,\tau')G(\tau',\tau) \big]$. To go further, one needs to write and solve the equation of motion for the two-particle Green functions for the QD, which is an ambitious task. However, for a non-interacting system, the QD two-particle Green functions can be decouple into a product of two QD one-particle Green functions. We get
\begin{eqnarray}\label{noise_tau_final}
&&\mathcal{S}_{\alpha\beta}(\tau,\tau')=\delta_{\alpha\beta}\mathcal{\tilde S}_{\alpha}(\tau,\tau')
+\frac{e^2}{\hbar^2}  \sum_{k\in\alpha,k'\in \beta} \vert V_k V_{k'} \vert^2 
\nonumber \\
&&
\times\iint d\tau_1 d\tau_2\Big[-g_k(\tau_1,\tau)g_{k'}(\tau_2,\tau')G(\tau,\tau_2)G(\tau',\tau_1)\nonumber \\
&&
 +g_k(\tau_2,\tau)g_{k'}(\tau',\tau_1) G(\tau,\tau')G(\tau_1,\tau_2)\nonumber \\
&& +g_k(\tau,\tau_1)g_{k'}(\tau_2,\tau') G(\tau',\tau)G(\tau_1,\tau_2)\nonumber \\ 
&&
 -g_k(\tau,\tau_1)g_{k'}(\tau',\tau_2) G(\tau_2,\tau)G(\tau_1,\tau') \Big] ~.
\end{eqnarray}
The different terms appearing on the right-hand side of Eq.~(\ref{noise_tau_final}) correspond to the connected part of the noise (the last four terms being the ones where the two integrals over $\tau_1$ and $\tau_2$ are intertwined). The last term $-\langle \hat I_\alpha \rangle\langle \hat I_\beta\rangle$ appearing on the right-hand side of Eq.~(\ref{noise_tau}) is no longer present since it is exactly canceled by the disconnected part of the noise (in which the two integrals over $\tau_1$ and $\tau_2$ can be done separately).

Next, we perform an analytical continuation of Eq.~(\ref{noise_tau_final}), we substitute the one-particle Green functions for the reservoirs and the QD by their expressions given in Ref.~\onlinecite{haug07} and we make a Fourier transform, assuming that all the Green functions are time translation invariant (see Appendix \ref{appendix1} for the details of the calculations). Moreover, the coupling strength between the QD and the reservoirs, $\Gamma_\alpha(\varepsilon)=2\pi |V_\alpha(\varepsilon)|^2\rho_\alpha(\varepsilon)$, with $\rho_\alpha(\varepsilon)$ the density of states in the reservoir $\alpha$ and  $V_\alpha(\varepsilon)=V_{k\in\alpha}$, is assumed to be energy independent (exact in the wide band limit). In addition, we consider symmetric coupling \cite{note2}, i.e., $\Gamma_L=\Gamma_R=\Gamma$ .

\subsection{Results and discussion}

Within these assumptions, we obtain the following expression for the NSFF noise
\begin{eqnarray}\label{NS_noise}
\mathcal{S}_{\alpha\beta}(\omega)=\frac{e^2}{h}\sum_{\gamma\delta}\int_{-\infty}^{\infty}d\varepsilon M_{\alpha\beta}^{\gamma\delta}(\varepsilon, \omega)f^e_\gamma(\varepsilon)f^h_\delta(\varepsilon-\hbar\omega)~,\nonumber \\
\end{eqnarray}
where $f^e_{\gamma}(\varepsilon)$ is the Fermi-Dirac distribution function, $f^h_{\delta}(\varepsilon)=1-f^e_{\delta}(\varepsilon)$, and where the matrix elements $M_{\alpha\beta}^{\gamma\delta}(\varepsilon, \omega)$ are listed in Table~I. They are expressed in terms of the transmission amplitude $t(\varepsilon)=i\Gamma G^r(\varepsilon)$, with $G^r(\varepsilon)$ the QD retarded Green function, and of the transmission coefficient $\mathcal{T}(\varepsilon)=t(\varepsilon)t^*(\varepsilon)$ through the barriers. The NSFF noise is composed of four contributions, each of which is given by the integral over the energy $\varepsilon$ of the matrix elements $M_{\alpha\beta}^{\gamma\delta}(\varepsilon,\omega)$ weighted by the product of the electron distribution function $f^e_{\gamma} (\varepsilon)$ and hole distribution function $f^h _{\delta} (\varepsilon-h\omega)$. When $\alpha=\beta$, Eq.~(\ref{NS_noise}) gives the expressions of the NSFF auto-correlators $\mathcal{S}_{LL}(\omega)$ and $\mathcal{S}_{RR}(\omega)$; 
when $\alpha\ne \beta$, it gives the expressions for the NSFF cross-correlators $\mathcal{S}_{LR}(\omega)$ 
and $\mathcal{S}_{RL}(\omega)$. The calculations presented here apply for a non-interacting QD. However, Eq.~(\ref{NS_noise}) holds even in the presence of Coulomb interactions in the QD provided that the QD two-particles Green function can be decoupled into a product of two QD one-particle Green functions as discussed above (equivalent to neglecting the vertex corrections). The effects of the interactions will then be entirely contained in the transmission amplitude $t(\varepsilon)$ and the coefficient $\mathcal{T}(\varepsilon)$.

Since $\mathcal{S}_{\alpha\beta}(\omega)=\mathcal{S}^*_{\beta\alpha}(\omega)$, the sum of $\mathcal{S}_{\alpha\beta}(\omega)$ over the indexes $\alpha$, $\beta$ of the two reservoirs, is a real quantity equal to
\begin{eqnarray}\label{double_sum}
\sum_{\alpha\beta}\mathcal{S}_{\alpha\beta}(\omega)&=&\frac{e^2}{h}\int_{-\infty}^{\infty}d\varepsilon|t(\varepsilon-\hbar\omega)-t(\varepsilon)|^2\nonumber\\
&&\times  \sum_{\gamma\delta}f^e_\gamma(\varepsilon)f^h_\delta(\varepsilon-\hbar\omega)~.
\end{eqnarray}
The above double sum is related to the charge fluctuations \cite{blanter00,rothstein09,clerk02} in the QD as shown in Appendix~\ref{appendix2}. Indeed, for an Anderson-type transmission amplitude, i.e., $t(\varepsilon)=i\Gamma/(\varepsilon-\varepsilon_0+i\Gamma)$, Eq.~(\ref{double_sum}) leads to $\sum_{\alpha\beta}\mathcal{S}_{\alpha\beta}(\omega)=\omega^2\mathcal{S}_\text{Q}(\omega)$, with ${S}_\text{Q}(\omega)=\int dt e^{-i\omega t}\langle \Delta\hat Q(t)\Delta\hat Q(0)\rangle$, the noise associated with the fluctuations of the charge $\hat Q=e d^\dag d$ in the QD. At zero-frequency or for an energy-independent transmission amplitude, $\sum_{\alpha\beta}\mathcal{S}_{\alpha\beta}(\omega)$ vanishes. This sum is non-zero only when $t(\varepsilon)$ acquires an energy dependence, as it is the case for example when the system is coupled to an electromagnetic environment~\cite{parmentier11}.

It is important to stress that our result for $\mathcal{S}_{LL}(\omega)$ differs from the expression of the NSFF noise given in Ref.~\onlinecite{nazarov09} in which the term $|t(\varepsilon)-t(\varepsilon-\hbar\omega)|^2$ is absent. However, when we symmetrize Eq.~(\ref{NS_noise}) with respect to the frequency, we get an expression that coincides exactly with the result obtained by B\"uttiker using scattering theory \cite{buttiker92} in which the emblematic term $|t(\varepsilon)-t(\varepsilon-\hbar\omega)|^2$ is also present. Moreover, at zero temperature, our result coincides with the expressions obtained by Hammer and Belzig in Ref.~\onlinecite{hammer11}, and at equilibrium, the fluctuation-dissipation theorem holds since we get $\mathcal{S}(\omega)=2\hbar\omega N(\omega)G(\omega)$, with $N(\omega)$, the Bose-Einstein distribution function and $G(\omega)$, the ac-conductance (the reservoir indexes can be removed in that limit). This confirms the validity of our calculations. We have also checked 
that by using scattering theory (non-interacting limit), we get an expression for the NSFF noise in terms of the matrix elements $M_{\alpha\beta}^{\gamma\delta}(\varepsilon,\omega)$, identical to that obtained in Eq.~(\ref{NS_noise}) using the NEGF technique.

\begin{widetext}

\begin{table}[h!]
\begin{center}
\begin{tabular}{|c||c|c|c|c|}
\hline
$M_{\alpha\beta}^{\gamma\delta}(\varepsilon,\omega)$& $\gamma=\delta=L$& $\gamma=\delta=R$&$\gamma=L$, $\delta=R$&$\gamma=R$, $\delta=L$\\ \hline\hline
$\alpha=L$&  $\mathcal{T}(\varepsilon)\mathcal{T}(\varepsilon-\hbar\omega)$& $\mathcal{T}(\varepsilon)\mathcal{T}(\varepsilon-\hbar\omega)$ & $\mathcal{T}(\varepsilon-\hbar\omega)[1-\mathcal{T}(\varepsilon)]$ &  $\mathcal{T}(\varepsilon)[1-\mathcal{T}(\varepsilon-\hbar\omega)]$\\
$\beta=L$&$+|t(\varepsilon)-t(\varepsilon-\hbar\omega)|^2$&&&\\
 \hline
$\alpha=R$& $\mathcal{T}(\varepsilon)\mathcal{T}(\varepsilon-\hbar\omega)$ &$\mathcal{T}(\varepsilon)\mathcal{T}(\varepsilon-\hbar\omega)$ & $\mathcal{T}(\varepsilon)[1-\mathcal{T}(\varepsilon-\hbar\omega)]$ & $\mathcal{T}(\varepsilon-\hbar\omega)[1-\mathcal{T}(\varepsilon)]$  \\
$\beta=R$&&$+|t(\varepsilon)-t(\varepsilon-\hbar\omega)|^2$&&\\
 \hline
$\alpha=L$& $t(\varepsilon)t^*(\varepsilon-\hbar\omega)$ & $t^*(\varepsilon)t(\varepsilon-\hbar\omega)$  & $t(\varepsilon)t(\varepsilon-\hbar\omega) [1-t^*(\varepsilon)]$& $t^*(\varepsilon)t^*(\varepsilon-\hbar\omega)[1-t(\varepsilon)]$\\
$\beta=R$&$\times [(1-t^*(\varepsilon))(1-t(\varepsilon-\hbar\omega))-1]$&$\times [(1-t(\varepsilon))(1-t^*(\varepsilon-\hbar\omega))-1]$&$\times[1-t^*(\varepsilon-\hbar\omega)]$&$\times [1-t(\varepsilon-\hbar\omega)]$\\
 \hline
$\alpha=R$& $t^*(\varepsilon)t(\varepsilon-\hbar\omega)$&$t(\varepsilon)t^*(\varepsilon-\hbar\omega)$   &$t^*(\varepsilon)t^*(\varepsilon-\hbar\omega)[1-t(\varepsilon)]$&
$t(\varepsilon)t(\varepsilon-\hbar\omega)[1-t^*(\varepsilon)]$\\
$\beta=L$&$\times[(1-t(\varepsilon))(1-t^*(\varepsilon-\hbar\omega))-1]$&$\times [(1-t^*(\varepsilon))(1-t(\varepsilon-\hbar\omega))-1]$&$\times [1-t(\varepsilon-\hbar\omega)]$ &$\times [1-t^*(\varepsilon-\hbar\omega)]$
\\
 \hline
\end{tabular}
\caption{Expressions of the matrix elements $M_{\alpha\beta}^{\gamma\delta}(\varepsilon, \omega)$ appearing in Eq.~(\ref{NS_noise}).}
\label{table1}
\end{center}
\end{table}

\end{widetext}


\section{Coherent superposition of scattering paths}

\begin{figure}
\begin{center}
\includegraphics[width=8cm]{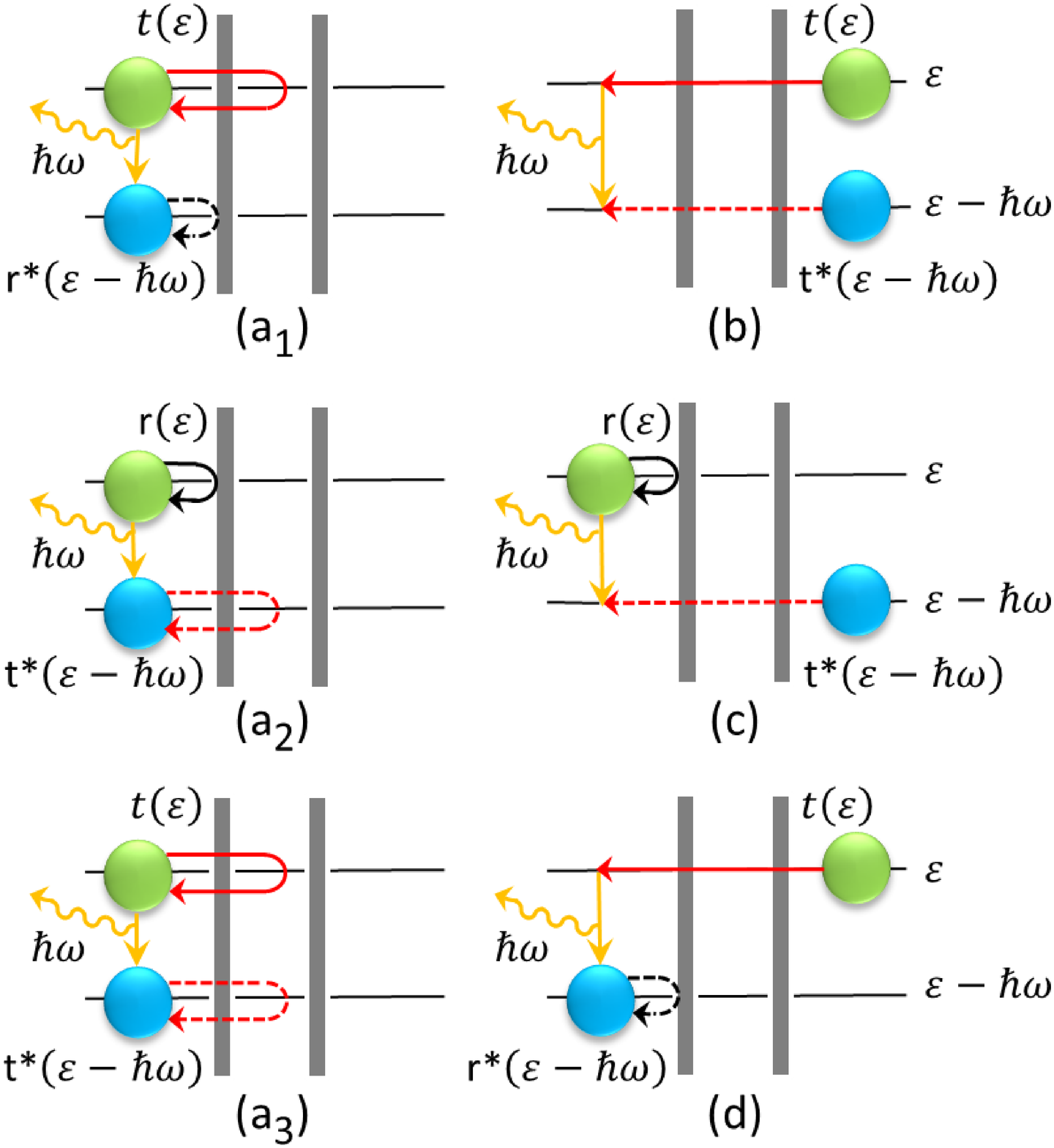}
\caption{Illustration of the processes that contribute to $\mathcal{S}_{LL}(\omega)$ with the transfer of energy $\hbar\omega$ (yellow wavy line) in the left reservoir. The green [blue] circle represents an electron [hole] with energy $\varepsilon$ [$\varepsilon-\hbar\omega$]. The solid [dashed] red arrow line represents a transmission process with probability amplitude $t(\varepsilon)$ [$t^*(\varepsilon-\hbar\omega)$]. The solid [dashed] black arrow line represents a reflection process with probability amplitude $r(\varepsilon)$ [$r^*(\varepsilon-\hbar\omega)$]. The coherent superposition of processes (a$_1$), (a$_2$), and (a$_3$) leads to the $M_{LL}^{LL}(\varepsilon, \omega)$ contribution,  process (b) to the $M_{LL}^{RR}(\varepsilon, \omega)$ contribution, process~(c) to the $M_{LL}^{LR}(\varepsilon, \omega)$ contribution, and process (d) to the $M_{LL}^{RL}(\varepsilon, \omega)$ contribution.}
\label{figure1}
\end{center} 
\end{figure}

Let us now proceed on to the second method that we develop to calculate the NSFF auto-correlators, i.e., $\mathcal{S}_{LL}(\omega)$ and $\mathcal{S}_{RR}(\omega)$, which correspond at positive frequency to the emission noises in the left and right reservoir, respectively~\cite{aguado00,deblock03}. It consists in identifying the physical processes involved in the generation of current noises. For this, one must go back to the definition of the noise given above, stating that it is the Fourier transform of a two-particle correlator, say, for instance, $\mathcal{S}_{LL}(t,0)=\langle \Delta \hat{I}_L(t) \Delta \hat{I}_L(0) \rangle$ for the auto-correlator in the left reservoir. From this and due to charge and energy conservation, one can see that $S_{LL}(\omega)$ corresponds to the transition probability from an initial state formed by creating a pair of one electron with energy $\varepsilon$ and one hole with energy $\varepsilon-\hbar\omega$ in either the left or the right reservoir, to a final state where the 
electron-hole pair is located in the left 
reservoir and then recombines emitting energy $\hbar\omega$ on the left reservoir side. Fig.~\ref{figure1} illustrates all the possible processes along which the system transits from such an initial state to that final state. Note that in order to contribute to the noise, the physical process must allow either the electron or the hole of the electron-hole pair to experience an excursion into the QD.

We first discuss the contributions to $\mathcal{S}_{LL}(\omega)$ when one starts from an initial state with an electron-hole pair located in the left reservoir, i.e., proportional to  $f^e_L(\varepsilon)f^h_L(\varepsilon-\hbar\omega)$. In this case, there exist three distinct processes allowing the electron or the hole to experience an excursion into the QD: in the first process, illustrated in Fig.~\ref{figure1}(a$_1$), the hole of the initial electron-hole pair is reflected by the left barrier while its electron partner moves back and forth between the left reservoir and the QD before emitting energy $\hbar\omega$ by recombining with its hole partner in the left reservoir. The corresponding transition probability amplitude is $t_1(\varepsilon, \omega)=t(\varepsilon)r^*(\varepsilon-\hbar\omega)$. In the second process, illustrated in Fig.~\ref{figure1}(a$_2$), the hole of the electron-hole pair moves back and forth between the left reservoir and the QD while its electron partner is reflected by the left 
barrier 
before emitting 
energy $\hbar\omega$ by recombining with the hole present in the left reservoir. Its amplitude reads $t_2(\varepsilon, \omega)=r(\varepsilon)t^*(\varepsilon-\hbar\omega)$. Finally in the third process, illustrated in Fig.~\ref{figure1}(a$_3$), both the hole and the electron of the electron-hole pair move back and forth between the left reservoir and the QD before emitting  energy $\hbar\omega$ by recombining together. Its amplitude is $t_3(\varepsilon, \omega)=t(\varepsilon)t^*(\varepsilon-\hbar\omega)$. These three processes lead to the following contribution to the noise: $\big|t_1(\varepsilon, \omega)+t_2(\varepsilon, \omega)+t_3(\varepsilon, \omega)\big|^2$
corresponding to the coherent superposition of the three transmission processes in question. The amplitudes rather than the probabilities have to be summed over the different processes since the wave function of the system is anti-symmetric under exchange of particles due to the Pauli principle. As a result, the processes in which electrons or holes are either transmitted or reflected are not distinguishable \cite{blanter00,martin05}. We have checked that $|t_1(\varepsilon, \omega)+t_2(\varepsilon, \omega)+t_3(\varepsilon, \omega)|^2$ is identical to $M_{LL}^{LL}(\varepsilon,\omega)=\mathcal{T}(\varepsilon)\mathcal{T}(\varepsilon-\hbar\omega)+|t(\varepsilon)-t(\varepsilon-\hbar\omega)|^2$ using the relations: $r(\varepsilon)=1-t(\varepsilon)$  and $t(\varepsilon)+t^*(\varepsilon)=2\mathcal{T}(\varepsilon)$.  These latter two relations hold for the non-interacting single-impurity model with symmetric barriers. The former relation arises from the matrix relation $\hat S(\varepsilon)=\hat 1-\hat t(\varepsilon)$ connecting the $\hat S$ matrix and the scattering matrix, whereas the latter relation corresponds to the generalized optical theorem \cite{newton66} resulting from the unitarity property of the $\hat S$ matrix -- holding when only elastic scattering of electrons occurs, as is the case here -- providing that the former relation is satisfied.

The physical processes involved in the three other contributions to the noise $\mathcal{S}_{LL}(\omega)$ can be identified in the same way. For the contribution proportional to  $f^e_R(\varepsilon)f^h_R(\varepsilon-\hbar\omega)$, one has to start from an initial state in which both the electron and the hole of the pair are in the right reservoir. There is a single process as illustrated in Fig.~\ref{figure1}(b). Both the electron and the hole have to move across the entire structure from the right to the left reservoir through the QD before recombining to emit energy $\hbar\omega$ in the left reservoir. The latter process leads to the following contribution to the noise:  $|t(\varepsilon)t^*(\varepsilon-\hbar\omega)|^2$, which is identical to $M_{LL}^{RR}(\varepsilon, \omega)=\mathcal{T}(\varepsilon)\mathcal{T}(\varepsilon-\hbar\omega)$. The same analysis can be carried out for the last two contributions proportional to  $f^e_L(\varepsilon)f^h_R(\varepsilon-\hbar\omega)$ and $f^e_R(\varepsilon)f^h_
L(\varepsilon-\hbar\omega)$
leading to the identification of processes (c) and (d) of Fig.~\ref{figure1} and to the expressions of $M_{LL}^{LR}(\varepsilon, \omega)$ and $M_{LL}^{RL}(\varepsilon, \omega)$.

Note that these processes describe the contributions to $\mathcal{S}_{LL}(\omega)$ up to any order in the coupling strength $\Gamma$. In the weak-coupling limit, the processes (a$_1)$, (a$_2)$, (c), and (d) are of order $\Gamma^2$, whereas the processes (a$_3)$ and (b) are of order $\Gamma^4$. Similar physical processes can be found to explain the contributions to $\mathcal{S}_{RR}(\omega)$. It is not possible, however, to find such a simple picture for the NSFF cross-correlators $\mathcal{S}_{LR}(\omega)$ and $\mathcal{S}_{RL}(\omega)$ since they are not real, and thus they are nonobservable quantities \cite{engel04}.


\section{Out-of-equilibrium noise spectrum and Fano factor}

To illustrate our results, we use the following exact form for the dot electron Green function in the non-interacting Anderson single-impurity model, $G^r(\varepsilon)=(\varepsilon-\varepsilon_0+i\Gamma)^{-1}$. This expression leads to a Lorentzian-type expression for the transmission amplitude $t(\varepsilon)=i\Gamma/(\varepsilon-\varepsilon_0+i\Gamma)$, where we recall that $\Gamma$ is the coupling strength between the QD and the reservoirs and $\varepsilon_0$ is the QD energy level. The auto-correlator and the cross-correlator spectrum, as well as the Fano factor defined as the ratio between the zero-frequency noise and the current, are quantities that are well discussed in the literature \cite{engel04,rothstein09,hammer11}. Here, we choose to focus on the following features: the conditions to observe distinct auto-correlators in the left and right reservoirs, and the profile of the differential Fano factor.

\subsection{NSFF noise in left and right reservoirs}

\begin{figure}[h!]
\begin{center}
\includegraphics[width=4cm]{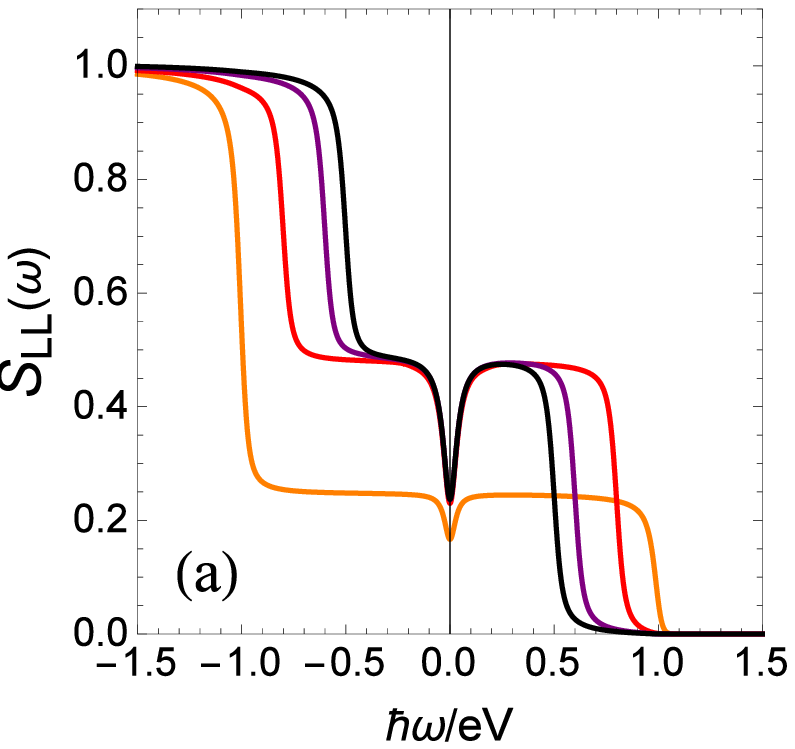}
\hspace{0.2cm}
\includegraphics[width=4cm]{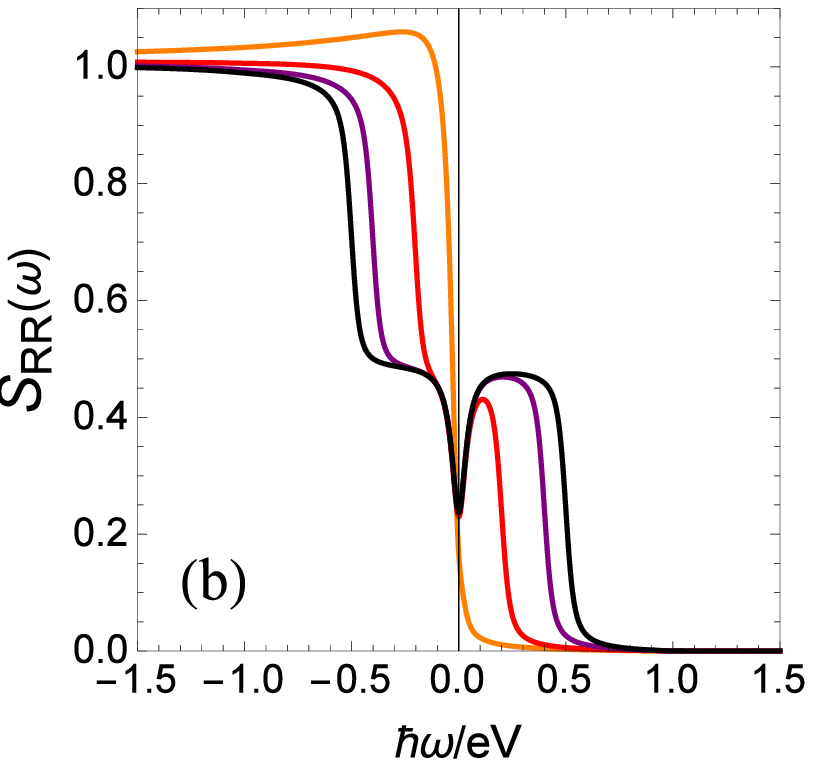}
\caption{Auto-correlators in units of $e^2\Gamma/\hbar$ in (a) the left reservoir and (b) the right reservoir at $k_BT/eV=0.01$, $\Gamma/eV=0.02$ and $\varepsilon_0=0$ for $\{\mu_L=eV,\mu_R=0\}$ (orange lines), $\{\mu_L=0.8eV,\mu_R=-0.2eV\}$ (red lines), $\{\mu_L=0.6eV,\mu_R=-0.4eV\}$ (purple lines) and  $\{\mu_L=0.5eV,\mu_R=-0.5eV\}$ (black lines). }
\label{figure2}
\end{center} 
\end{figure}

From Table~\ref{table1}, we see that the expressions for the auto-correlators in the left and the right reservoirs are distinct. For example, in the zero-temperature limit and at positive frequency and positive bias voltage, the auto-correlators are given by
\begin{eqnarray}
\mathcal{S}_{LL}(\omega)&=&\frac{e^2}{h}\int_{\mu_R+\hbar\omega}^{\mu_L}d\varepsilon \mathcal{T}(\varepsilon-\hbar\omega)\big[1-\mathcal{T}(\varepsilon)\big]~,
\end{eqnarray}
and
\begin{eqnarray}
\mathcal{S}_{RR}(\omega)&=&\frac{e^2}{h}\int_{\mu_R+\hbar\omega}^{\mu_L}d\varepsilon \mathcal{T}(\varepsilon)\big[1-\mathcal{T}(\varepsilon-\hbar\omega)\big]~,
\end{eqnarray}
which are non-equal when the frequency is non-zero and when the transmission coefficient is energy-dependent. Such a difference between the left and right auto-correlators is also reported in Refs.~[\onlinecite{lu13,rothstein14}]. Its origin is the distinct energy of the carriers contributing to $\mathcal{S}_{LL}(\omega)$ or to $\mathcal{S}_{RR}(\omega)$ when transferred through the junction ($\varepsilon-\hbar\omega$ versus $\varepsilon$). There is an additional condition that appears when one plots the noise spectrum: the potential profile through the junction has to be non symmetric in the sense that the dot energy should not stand in the middle of the potential barrier, i.e. $\varepsilon_0\ne(\mu_L+\mu_R)/2$. Indeed, in Fig.~\ref{figure2} (in which we take $\varepsilon_0=0$), we see that when $\mu_L+\mu_R\ne 0$, the auto-correlators in the left and right reservoirs are different (compare orange, red and purple lines) and that it is only when  $\mu_L+\mu_R=0$, that these two quantities coincide (compare black lines). Thus, $\mathcal{S}_{LL}(\omega)$ differs from $\mathcal{S}_{RR}(\omega)$ provided that the following three conditions are fulfilled: non-zero frequency, energy dependent transmission and non-symmetric potential profile.

As expected in the zero temperature limit, the emission noises $\mathcal{S}_{LL}(\omega)$ and $\mathcal{S}_{RR}(\omega)$ (at positive frequency) drop to zero for frequency higher than voltage: the system can not supply energy larger than the energy provided to it (here the voltage). The absorption noise (at negative frequency) has no such limitation. We observe that for $\hbar\omega<-eV$, $\mathcal{S}_{LL}(\omega)$ and $\mathcal{S}_{RR}(\omega)$ converge to the same value (equal to $e^2\Gamma/\hbar$), even when the above conditions meet. Finally,  $\mathcal{S}_{LL}(\omega)$ and $\mathcal{S}_{RR}(\omega)$ differ from each other in the interval $\hbar\omega\in[-eV,0[\;\cup\;]0,eV]$ provided that $\mathcal{T}(\varepsilon)$ is energy-dependent and $\varepsilon_0\ne(\mu_L+\mu_R)/2$. In any other situations, the auto-correlators in the left and right reservoirs coincide.


\subsection{Differential Fano factor}

\begin{figure}[h!]
\begin{center}
\includegraphics[width=9cm]{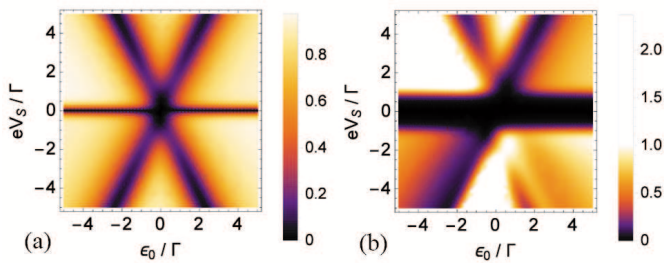}
\caption{Differential Fano factor $F(V_S,\omega)$ as a function of gate energy~$\varepsilon_0$ and bias voltage $V_S$ (both normalized by $\Gamma$) at $k_BT/\Gamma=0.1$ for two values of the frequency : (a) $\hbar\omega_1/\Gamma=0.1$ (low frequency) and (b) $\hbar\omega_2/\Gamma=1$ (high frequency).}
\label{figure3}
\end{center} 
\end{figure}

The differential Fano factor was introduced in recent experimental works~\cite{basset12} and defined as  $F(V_S,\omega)=\big|[d\mathcal{S}_{LL}/dV]_{V_S}/[ed\langle \hat I_L\rangle/dV]_{V_S-\hbar\omega/e}\big|$. It corresponds to the ratio of the derivative of the noise with bias voltage at $V=V_S$, to the differential conductance at $V=V_S-\hbar\omega/e$, where $\omega$ is the frequency at which the noise is measured. We plot $F(V_S,\omega)$ in Fig.~\ref{figure3} as a function of gate energy $\varepsilon_0$ and bias voltage $V_S$ at two values of the frequency, $\hbar\omega_1 /\Gamma=0.1$ and $\hbar\omega_2 /\Gamma=1$. In agreement with the fact that the system cannot emit energy larger than the supplied energy $eV_S$, we observe that $F(V_S,\omega)$ is strongly reduced for bias voltages $V_S$ smaller than the frequency in absolute 
value (see the black horizontal band in both graphs of Fig.~\ref{figure3}). At low frequency (Fig.~\ref{figure3}(a)), $F(V_S,\omega)$ is always smaller than 1, meaning that the noise is sub-Poissonian. At higher frequency (Fig.~\ref{figure3}(b)), $F(V_S,\omega)$ takes values higher than 1 in some regions, meaning that the noise is super-Poissonian with an upper limit for $F(V_S,\omega)$ equal to 2, in agreement with Ref.~\onlinecite{hammer11}. Such an increase of $F(V_S,\omega)$ with increasing frequency has been measured in a carbon nanotube QD when placed in the Kondo regime~\cite{basset12}. Even if the analysis carried out here is not intended to describe the Kondo effect, it is worth noticing that this increase of $F(V_S,\omega)$ with increasing frequency constitutes a general trend. At both low and high frequencies, $F(V_S,\omega)$ is reduced in the bands surrounding the first bisectors, $\varepsilon_0=\pm eV_S/2$ in the case of a symmetric profile of the chemical potentials on either sides of the QD, i.
e. $\mu_{L,R}=\pm eV_S/2$. These bands correspond to a maximal 
conductance with a transmission close to 1. The results reported in the two graphs show similarities with what has been observed experimentally in Ref.~\onlinecite{basset12} with a characteristic pattern composed of six areas of the non-zero differential Fano factor separated by narrow bands in which the differential Fano factor is strongly reduced. Experimentally in the presence of the Kondo effect, the bands surrounding the bisectors are no longer straight but become curved due to the effect of the interactions.

\section{Conclusion}

In summary, we have derived the expression of the NSFF noises in a single-level QD connected to reservoirs using the NEGF technique. Both the auto- and cross-correlators are expressed in terms of the transmission amplitude $t(\varepsilon)$ and coefficient $\mathcal{T}(\varepsilon)$. We have identified the physical processes at the origin of the contributions to the auto-correlators. We have shown that when the electron-hole pair is initially present in the reservoir emitting the energy, the contributions to the noise can be interpreted in terms of a coherent superposition of three distinct processes. On the contrary, when the electron and/or the hole of the pair is initially in the non-emitting reservoir, only one type of physical process exists and contributes to the noise. We have also shown that for a non-interacting QD, the differential Fano factor as a function of gate and bias voltages presents interesting features reminiscent of the experimental measurements. The methods 
presented and tested here could be extended to treat many situations involving Coulomb interactions and other transmission amplitudes and coefficients, such as multiple channel systems, QDs with multiple energy levels, or quantum point contacts embedded in an electromagnetic environment.

\acknowledgments

We would like to thank Y.M.~Blanter, R.~Deblock, M.~Guigou, T.~Martin, F.~Michelini, Y.V.~Nazarov and F.~Portier for valuable discussions. For financial support, the authors acknowledge the Indo-French Centre for the Promotion of Advanced Research (IFCPAR) under Research Project No.4704-02.

* Also at the Centre National de la Recherche Scientifique (CNRS).


\appendix

\section{Current noise}\label{appendix1}

From Eq.~(\ref{noise_tau_final}), we have $\mathcal{S}_{\alpha\beta}(\tau,\tau')=\sum_{i=1}^5\mathcal{C}_{\alpha\beta}^{(i)}(\tau,\tau')$ with
\begin{eqnarray}
\mathcal{C}_{\alpha\beta}^{(1)}(\tau,\tau')&=&\frac{e^2}{\hbar^2}\sum_{k\in \alpha} \vert V_k \vert^2 \Big[g_k(\tau',\tau)G(\tau,\tau')\nonumber\\
&&+g_k(\tau,\tau')G(\tau',\tau) \Big]\delta_{\alpha\beta}~,
\end{eqnarray}
\begin{eqnarray}
&&\mathcal{C}_{\alpha\beta}^{(2)}(\tau,\tau')=-\frac{e^2}{\hbar^2}\sum_{k\in\alpha,k'\in\beta} \vert V_k V_{k'} \vert^2 \iint d\tau_1 d\tau_2 \nonumber\\
&&\times G(\tau,\tau_2)g_{k'}(\tau_2,\tau')G(\tau',\tau_1)g_k(\tau_1,\tau)~,
\end{eqnarray}
\begin{eqnarray}
&&\mathcal{C}_{\alpha\beta}^{(3)}(\tau,\tau')=\frac{e^2}{\hbar^2} \sum_{k\in\alpha,k'\in\beta} \vert V_k V_{k'} \vert^2 \iint d\tau_1 d\tau_2\nonumber\\
&&\times G(\tau,\tau') g_{k'}(\tau',\tau_1)G(\tau_1,\tau_2)g_k(\tau_2,\tau)~,
\end{eqnarray}
\begin{eqnarray}
&&\mathcal{C}_{\alpha\beta}^{(4)}(\tau,\tau')=\frac{e^2}{\hbar^2}\sum_{k\in\alpha, k'\in \beta} \vert V_k V_{k'} \vert^2 \iint d\tau_1 d\tau_2\nonumber\\
&&\times  g_{k}(\tau,\tau_1)  G(\tau_1,\tau_2)g_{k'}(\tau_2,\tau')G(\tau',\tau)~,
\end{eqnarray}
\begin{eqnarray}
&&\mathcal{C}_{\alpha\beta}^{(5)}(\tau,\tau')=-\frac{e^2}{\hbar^2} \sum_{k\in\alpha, k'\in \beta} \vert V_k V_{k'} \vert^2 \iint d\tau_1 d\tau_2 \nonumber\\
&&\times  g_k(\tau,\tau_1)G(\tau_1,\tau')   g_{k'}(\tau',\tau_2)G(\tau_2,\tau)~.
\end{eqnarray}
Performing the analytical continuation \cite{haug07} of the previous equations, we obtain  $\mathcal{S}_{\alpha\beta}(t,t')=\sum_{i=1}^5\mathcal{C}_{\alpha\beta}^{(i)}(t,t')$ with
\begin{eqnarray}
\mathcal{C}_{\alpha\beta}^{(1)}(t,t')&=&\frac{e^2}{\hbar^2}\sum_{k\in \alpha} \vert V_k \vert^2 \Big[g_k^{<}(t',t)G^>(t,t')\nonumber\\
&&+g_k^{>}(t,t')G^<(t',t) \Big]\delta_{\alpha\beta}
~,
\end{eqnarray}
\begin{eqnarray}
&&\mathcal{C}_{\alpha\beta}^{(2)}(t,t')=-\frac{e^2}{\hbar^2}
 \sum_{k\in\alpha,k'\in\beta} \vert V_k  V_{k'} \vert^2 \iint dt_1 dt_2\nonumber \\
&&\times \Big[ G^{r}(t',t_1)g_k^{<}(t_1,t)+G^{<}(t',t_1)g_k^{a}(t_1,t) \Big] \nonumber \\
 &&\times \Big[ G^{>}(t,t_2)g_{k'}^{a}(t_2,t')+G^{r}(t,t_2)g_{k'}^{>}(t_2,t') \Big]~,
\end{eqnarray}
\begin{eqnarray}
\mathcal{C}_{\alpha\beta}^{(3)}(t,t')&=&\frac{e^2}{\hbar^2}\sum_{k\in\alpha,k'\in\beta} \vert V_k V_{k'} \vert^2 G^>(t,t') \iint dt_1 dt_2 \nonumber \\ 
 &&\times \Big[ g_{k'}^{r}(t',t_1)G^{r}(t_1,t_2)g_k^{<}(t_2,t)\nonumber \\
&&+g_{k'}^{r}(t',t_1)G^{<}(t_1,t_2)g_k^{a}(t_2,t) \nonumber \\
&&+g_{k'}^{<}(t',t_1)G^{a}(t_1,t_2)g_k^{a}(t_2,t) \Big]~,
\end{eqnarray}
\begin{eqnarray}
\mathcal{C}_{\alpha\beta}^{(4)}(t,t')&=&\frac{e^2}{\hbar^2}\sum_{k\in\alpha,k'\in\beta} \vert V_k  V_{k'} \vert^2 G^<(t',t) \iint dt_1 dt_2 \nonumber \\ 
&&\times \Big[ g_{k}^{>}(t,t_1)G^{a}(t_1,t_2)g_{k'}^{a}(t_2,t')\nonumber \\
&&+g_{k}^{r}(t,t_1)G^{>}(t_1,t_2)g_{k'}^{a}(t_2,t')\nonumber \\
&&+g_{k}^{r}(t,t_1)G^{r}(t_1,t_2)g_{k'}^{>}(t_2,t') \Big]~,
\end{eqnarray}
\begin{eqnarray}
&&\mathcal{C}_{\alpha\beta}^{(5)}(t,t')=-\frac{e^2}{\hbar^2}\sum_{k\in\alpha,k'\in\beta} \vert V_k V_{k'} \vert^2 \iint dt_1dt_2 \nonumber \\
&&\times\left[ g_k^{>}(t,t_1)G^{a}(t_1,t')+g_k^{r}(t,t_1)G^{>}(t_1,t') \right] \nonumber \\
 &&\times\left[ g_{k'}^{r}(t',t_2)G^{<}(t_2,t')+g_{k'}^{<}(t',t_2)G^{a}(t_2,t') \right]~.\nonumber \\
\end{eqnarray}
Considering the Fourier transform of the Green functions in the latter expressions in the case of a time translation invariance (steady state), we get $\mathcal{S}_{\alpha\beta}(\omega)=\sum_{i=1}^5\mathcal{C}_{\alpha\beta}^{(i)}(\omega)$ with
\begin{eqnarray}
\mathcal{C}_{\alpha\beta}^{(1)}(\omega)&=&\frac{e^2}{h}\sum_{k\in \alpha} \vert V_k \vert^2 \int_{-\infty}^{\infty} d\varepsilon \Big[ g_k^{<}(\varepsilon)G^>(\varepsilon-\hbar\omega)\nonumber\\
&&+g_k^{>}(\varepsilon-\hbar\omega)G^<(\varepsilon) \Big]\delta_{\alpha\beta}~,
\end{eqnarray}
\begin{eqnarray}
&&\mathcal{C}_{\alpha\beta}^{(2)}(\omega)=-\frac{e^2}{h}\sum_{k\in\alpha,k'\in \beta} \vert V_k V_{k'} \vert^2 \nonumber\\
&&\times\int_{-\infty}^{\infty} d\varepsilon\Big[G^{r}(\varepsilon)g_k^{<}(\varepsilon)G^{r}(\varepsilon-\hbar\omega)g_{k'}^{>}(\varepsilon-\hbar\omega)\nonumber\\
&&+G^{r}(\varepsilon)g_k^{<}(\varepsilon)G^{>}(\varepsilon-\hbar\omega)g_{k'}^{a}(\varepsilon-\hbar\omega)\nonumber\\
&&+G^{<}(\varepsilon)g_k^{a}(\varepsilon)G^{r}(\varepsilon-\hbar\omega)g_{k'}^{>}(\varepsilon-\hbar\omega)\nonumber\\
&&+G^{<}(\varepsilon)g_k^{a}(\varepsilon)G^{>}(\varepsilon-\hbar\omega)g_{k'}^{a}(\varepsilon-\hbar\omega)\Big]~,
\end{eqnarray}
\begin{eqnarray}
&&\mathcal{C}_{\alpha\beta}^{(3)}(\omega)=\frac{e^2}{h}\sum_{k\in\alpha,k'\in \beta} \vert V_k  V_{k'} \vert^2\nonumber\\
&&\times\int_{-\infty}^{\infty} d\varepsilon \Big[G^>(\varepsilon-\hbar\omega)g_{k'}^{r}(\varepsilon)G^{r}(\varepsilon)g_k^{<}(\varepsilon)\nonumber\\
&&+G^>(\varepsilon-\hbar\omega)g_{k'}^{r}(\varepsilon)G^{<}(\varepsilon)g_k^{a}(\varepsilon)\nonumber\\
&&+G^>(\varepsilon-\hbar\omega)g_{k'}^{<}(\varepsilon)G^{a}(\varepsilon)g_k^{a}(\varepsilon)\Big]~,
\end{eqnarray}
\begin{eqnarray}
&&\mathcal{C}_{\alpha\beta}^{(4)}(\omega)=\frac{e^2}{h} \sum_{k\in\alpha, k'\in \beta} \vert V_k V_{k'} \vert^2
\nonumber\\
&&\times\int_{-\infty}^{\infty} d\varepsilon 
\Big[G^<(\varepsilon)g_{k}^{>}(\varepsilon-\hbar\omega)G^{a}(\varepsilon-\hbar\omega)g_{k'}^{a}(\varepsilon-\hbar\omega)\nonumber\\
&&+G^<(\varepsilon)g_{k}^{r}(\varepsilon-\hbar\omega)G^{>}(\varepsilon-\hbar\omega)g_{k'}^{a}(\varepsilon-\hbar\omega)\nonumber\\
&&+G^<(\varepsilon)g_{k}^{r}(\varepsilon-\hbar\omega)G^{r}(\varepsilon-\hbar\omega)g_{k'}^{>}(\varepsilon-\hbar\omega)\Big]
~,
\end{eqnarray}
\begin{eqnarray}
&&\mathcal{C}_{\alpha\beta}^{(5)}(\omega)=-\frac{e^2}{h} \sum_{k\in\alpha,k'\in \beta} \vert V_k V_{k'} \vert^2\nonumber\\
&&\times\int_{-\infty}^{\infty} d\varepsilon\Big[ G^{a}(\varepsilon-\hbar\omega)g_k^{>}(\varepsilon-\hbar\omega)G^{<}(\varepsilon)g_{k'}^{r}(\varepsilon)\nonumber\\
&&+G^{a}(\varepsilon-\hbar\omega)g_k^{>}(\varepsilon-\hbar\omega)G^{a}(\varepsilon)g_{k'}^{<}(\varepsilon)\nonumber\\
&&+G^{>}(\varepsilon-\hbar\omega)g_k^{r}(\varepsilon-\hbar\omega)G^{<}(\varepsilon)g_{k'}^{r}(\varepsilon)\nonumber\\
&&+ G^{>}(\varepsilon-\hbar\omega)g_k^{r}(\varepsilon-\hbar\omega)G^{a}(\varepsilon)g_{k'}^{<}(\varepsilon)\Big]~.
\end{eqnarray}
In the expressions of each contribution $\mathcal{C}_{\alpha\beta}^{(i)}(\omega)$ to the noise, we report the bar Green functions of the reservoir
\begin{eqnarray}
g_{k\in\alpha}^{<}(\varepsilon)&=&2i\pi f^e_\alpha(\varepsilon) \delta(\varepsilon-\varepsilon_k)~,\\
g_{k\in\alpha}^{>}(\varepsilon)&=&-2i\pi f^h_\alpha(\varepsilon)\delta(\varepsilon-\varepsilon_k)~,\\
g_{k\in\alpha}^{r}(\varepsilon)&=&[\varepsilon-\varepsilon_k+i0^+]^{-1}~,\\
g_{k\in\alpha}^{a}(\varepsilon)&=&[\varepsilon-\varepsilon_k-i0^+]^{-1}~,
\end{eqnarray}
where $f^e_\alpha(\varepsilon)$ is the Fermi-Dirac distribution function associated with the reservoir $\alpha$, and $f^h_\alpha(\varepsilon)=1-f^e_\alpha(\varepsilon)$.
In the wide-band approximation limit, where $\Gamma_\alpha(\varepsilon)=2\pi\rho_\alpha(\varepsilon)\vert V(\varepsilon) \vert^2$ is independent of energy, we obtain the following expression for the NSFF noise
\begin{eqnarray}\label{noise_app}
&&\mathcal{S}_{\alpha\beta}(\omega)=\frac{e^2}{h}\Gamma_\alpha\delta_{\alpha\beta}\int_{-\infty}^{\infty} d\varepsilon \Big[
-if^h_\alpha(\varepsilon-\hbar\omega)G^<(\varepsilon)\nonumber \\
&&+if^e_\alpha(\varepsilon)\big[G^r(\varepsilon-\hbar\omega)-G^a(\varepsilon-\hbar\omega)
+G^<(\varepsilon-\hbar\omega)\big]\Big] \nonumber \\
&&+\frac{e^2}{h}\Gamma_\alpha \Gamma_\beta\int_{-\infty}^{\infty} d\varepsilon \Big[G^<(\varepsilon)G^>(\varepsilon-\hbar\omega)\nonumber \\
&&-f^e_\alpha(\varepsilon)f^h_\beta(\varepsilon-\hbar\omega) G^r(\varepsilon)G^r(\varepsilon-\hbar\omega)\nonumber \\
&&-f^e_\beta(\varepsilon)f^h_\alpha(\varepsilon-\hbar\omega)G^a(\varepsilon) G^a(\varepsilon-\hbar\omega) \nonumber \\
&&+ \big[f^e_\alpha(\varepsilon)G^r(\varepsilon)-f^e_\beta(\varepsilon)G^a(\varepsilon)\big]G^>(\varepsilon-\hbar\omega) \nonumber \\
&&+\big[f^h_\alpha(\varepsilon-\hbar\omega)G^a(\varepsilon-\hbar\omega)\nonumber \\
&&-f^h_\beta(\varepsilon-\hbar\omega) G^r(\varepsilon-\hbar\omega)\big]G^<(\varepsilon)\Big]~. 
\end{eqnarray}
For a non-interacting QD, we have \cite{haug07}
\begin{eqnarray}
&&G^r(\varepsilon)-G^a(\varepsilon)=-iG^r(\varepsilon)\big[\Gamma_L+\Gamma_R\big]G^a(\varepsilon)~,\nonumber \\~\\
&&G^<(\varepsilon)=iG^r(\varepsilon)\big[\Gamma_Lf^e_L(\varepsilon)+\Gamma_Rf^e_R(\varepsilon)\big]G^a(\varepsilon)~. \nonumber \\
\end{eqnarray}
Using $i\Gamma G^r(\varepsilon)=t(\varepsilon)$ and $G^>(\varepsilon)=G^r(\varepsilon)-G^a(\varepsilon)+G^<(\varepsilon)$, and considering symmetrical barriers, $\Gamma_L=\Gamma_R=\Gamma$, we get
\begin{eqnarray}
&&G^r(\varepsilon)-G^a(\varepsilon)=-\frac{2it(\varepsilon)t^\ast(\varepsilon)}{\Gamma}~,\\\label{AA1}
&&G^<(\varepsilon)=\frac{it(\varepsilon)t^\ast(\varepsilon)}{\Gamma}\big[f^e_L(\varepsilon)+f^e_R(\varepsilon)\big]~,\\\label{AA2}
&&G^>(\varepsilon)=-\frac{it(\varepsilon)t^\ast(\varepsilon)}{\Gamma}\big[f^h_L(\varepsilon)+f^h_R(\varepsilon)\big]~.
\end{eqnarray}
Incorporating the latter expressions into Eq.~(\ref{noise_app}), we get Eq.~(\ref{NS_noise}) with matrix elements $M_{\alpha\beta}^{\gamma\delta}(\varepsilon,\omega)$ given in Table~1.

\section{Charge noise}\label{appendix2}

The charge noise is defined as the Fourier transform of the charge fluctuations on the QD
\begin{eqnarray}
 \mathcal{S}_\text{Q}(\omega)=\int_{-\infty}^{\infty} dt e^{-i\omega t}\langle \Delta\hat Q(t)\Delta\hat Q(0)\rangle~,
\end{eqnarray}
where $\Delta\hat Q(t)=\hat Q(t)-\langle \hat Q\rangle$, with $\hat Q(t)=e\hat N(t)=ed^\dag(t) d(t)$. Performing the decoupling of the QD two-particles Green function, we get
\begin{eqnarray}
\langle \Delta\hat Q(t)\Delta\hat Q(0)\rangle=\langle d^\dag(t) d(0)\rangle \langle d(t) d^\dag(0) \rangle~,
\end{eqnarray}
which leads to
\begin{eqnarray}
 \mathcal{S}_\text{Q}(\omega)&=&e^2\int_{-\infty}^{\infty} dt e^{-i\omega t}G^<(0,t)G^>(t,0)\nonumber\\
&=&\frac{e^2}{h}\int_{-\infty}^{\infty} d\varepsilon G^<(\varepsilon)G^>(\varepsilon-\hbar\omega)~.
\end{eqnarray}
Using Eqs.~(\ref{AA1}) and (\ref{AA2}) and introducing the transmission amplitude $t(\varepsilon)=i\Gamma G^r(\varepsilon)$, we finally obtain
\begin{eqnarray}\label{SQ}
 \mathcal{S}_\text{Q}(\omega)&=&\frac{e^2}{h\Gamma^2}\int_{-\infty}^{\infty}d\varepsilon \mathcal{T}(\varepsilon)\mathcal{T}(\varepsilon-\hbar\omega)\nonumber\\
&&\times\sum_{\gamma\delta}f^e_\gamma(\varepsilon)f^h_\delta(\varepsilon-\hbar\omega)~.
\end{eqnarray}
In the case of an Anderson-type transmission amplitude, $t(\varepsilon)=i\Gamma/(\varepsilon-\varepsilon_0+i\Gamma)$, we have
\begin{eqnarray}
|t(\varepsilon)-t(\varepsilon-\hbar\omega)|^2=\frac{\omega^2}{\Gamma^2} \mathcal{T}(\varepsilon)\mathcal{T}(\varepsilon-\hbar\omega)~,
\end{eqnarray}
which leads when one compare Eq.~(\ref{SQ}) to Eq.~(\ref{double_sum}) to the equality
\begin{eqnarray}
\sum_{\alpha\beta}\mathcal{S}_{\alpha\beta}(\omega)=\omega^2   \mathcal{S}_\text{Q}(\omega)~.
\end{eqnarray}



\begin{thebibliography}{20}
\bibitem{note1} Note that Eq.~(\ref{correlator}) corresponds to the nonsymmetrized current correlator whereas the expression given in Ref.~\onlinecite{haug07} corresponds to the symmetrized one.
\bibitem{note2} Note that when one relaxes the assumption on symmetrical barriers, the expression of the noise in terms of the transmission amplitude is much more complicated since one has to consider all the possible writings, introducing all the possible transmission amplitudes $t_{LL}(\varepsilon)$,  $t_{RR}(\varepsilon)$,  $t_{LR}(\varepsilon)$ and $t_{RL}(\varepsilon)$.
\end{thebibliography}
\end{document}